# Complexity and chaotic behavior of the U.S. rivers and estimation of their prediction horizon

Dragutin T. Mihailović[a,*], Slavica Malinović-Milićević[b], Jeongwoo Han[c], Vijay P. Singh[d]

[a]Department of Physics, Faculty of Sciences, University of Novi Sad, Novi Sad, Serbia; guto@df.uns.ac.rs

[b]Geographical Institute "Jovan Cvijić", Serbian Academy of Sciences and Arts, Belgrade, Serbia; s.malinovic-milicevic@gi.sanu.ac.rs

[c]Department of Biological and Agricultural Engineering, Texas A&M University, College Station, TX, USA; han820124@ tamu.edu

[d]Department of Biological and Agricultural Engineering and Zachry Department of Civil & Environmental Engineering, Texas A&M University, College Station, TX 77843-2117, USA; vsingh@tamu.edu

*Correspondence: guto@df.uns.ac.rs; tel.: +38121458449

**Abstract**

A streamflow time series encompasses a large amount of hidden information and reliable prediction of its behavior in the future remains a challenge. It seems that the use of information measures can significantly contribute to determining the time horizon of rivers and improving predictability. Using the Kolmogorov complexity (KC) and its derivatives (KC spectrum and its highest value), and Lyapunov exponent (LE), it has previously been shown that the degree of streamflow predictability depends on human activities, environmental factors, and natural characteristics. This paper applied the KC and LE measures to investigate the randomness and chaotic behavior of monthly streamflow of 1879 rivers from the United States for a period of 1950–2015 and evaluated their time horizons via the Lyapunov and Kolmogorov time (LT and KT, respectively).

# 1. Introduction

*1.1 Considering the turbulent and chaotic behavior of rivers*

When we look at a wide river in the lowland, it seems calm and not much turbulent. However, it is only the impression caused by our perception. Birnir (2008) theoretically showed the solutions that describe turbulent flow in rivers and also included an invariant measure for describing the statistical properties of one-dimensional turbulence. Reynolds number is often used to characterize turbulent flow in rivers and streams. This number for rivers ($Re_{riv}$) is calculated as $Re_{riv} = \overline{DV}/\nu$, where $\overline{D}$ is the average depth of flow, $\overline{V}$ is the average velocity, and $\nu$ the kinematic viscosity. For streams and rivers, $Re_{riv}$ is typically large ( $Re_{riv} = 10^5 - 10^6$) (Dingman, 1984). The turbulence has much more degrees of freedom than flows in a chaotic mode. On the contrary, all chaotic flows are not necessarily turbulent. According to Li (2014), the relationship between turbulence and chaos can be described as follows: "when the Reynolds number is large, violent fully developed turbulence is due to 'rough dependence on initial data' rather than chaos which is caused by 'sensitive dependence on initial data'; when the Reynolds number is moderate, turbulence is due to chaos." Pursuing this relationship, rivers are *par excellence* complex systems that can have a high level of complexity and chaotic behavior. Precisely, chaos has a very accurate mathematical definition, while turbulence is a property of fluid flow, that has no accurate *mathematical* definition. In rivers, spatial and temporal irregular fluctuations, small as well as large, co-occur as three-dimensional eddies. It is difficult to prove whether these are stochastic or chaotically deterministic. Therefore, turbulence can be (i) one example of the physical manifestation of deterministic chaos, or (ii) a stochastic, non-chaotic, manifestation of the solution to the nonlinear fluid flow problem at *high* Reynolds numbers. The phenomenon we observe in river flow systems emerges from an underlying disorder and we embrace the noise and uncertainty as an essential step on the road toward *predictability*. The predictability of river streamflow usually refers to (1) the time evolution of the system from which we can obtain information and (2) the content of obtained information. Thus, our attention is mostly on a macroscopic model that predicts the state of the system for a longer period of time and larger spatial scale. Because of the complex nature of rivers, it is difficult to estimate their *prediction horizon* (Mihailović et al., 2022). There are some existing methods for its estimation (Regonda et al., 2013), but all of them have at least one drawback that does not allow reaching a reliable estimation.

*1.2 Studying streamflow complexity*

Understanding the dynamic behavior of rivers which is affected by several factors is a key issue in hydrology. Streamflow is affected by (i) physical factors that include the incline gradient of the river, water viscosity, elevation, and properties of the surrounding terrain; (ii) geophysical factors involving the geographical location, weather, and climatic change; and finally, it is significantly affected by (iii) human activities (including building, river training works, damming, dredging, deforestation, and pollution). The question arises here as to how the study of river flow complexity can help unravel the effects of these factors. In the context of complexity, as it will be discussed in this paper, we will cite (i) the paper by Puente and Sivakumar (2007) in which stream flow complexity is considered within the general geophysical complexity and (ii) the paper by Krasovskaia (1997) is a representative of a large group of papers in which streamflow complexity is described through the difference between time series by offering a correlational explanation.

Information about river streamflow complexity is most reliable if it is computed from time series by applying some information measures. It seems that algorithmic complexity can be a good choice, although this measure has not yet found its niche in hydrology except in a few papers (Sen, 2009; Mihailović et al., 2014; Mihailović et al., 2017). The use of hydrological models in studying streamflow complexity is not so promising, since with them it is not possible to model perhaps the most discriminating property of a complex system - complexity. With this in mind, we applied the Kolmogorov complexity (KC) to monthly streamflow time series. It is noted that streamflow complexity can be studied by the Aksentijevic-Gibson complexity as a tool for the analysis of hydrological data that holds the promise of uncovering patterns in the data that cannot be captured by KC and other complexity measures (Aksentijevic et al., 2021).

*1.3 Prediction horizon of rivers*

In mathematics, there exists a characteristic timescale well known as the Lyapunov time, also called *prediction horizon* (which is expressed in the units of the recorded series) defined as the inverse of the largest Lyapunov exponent of the considered time series. It is a period after which a dynamical system becomes unpredictable and enters a chaotic state, so it indicates the limits of predictability. Estimation of the Lyapunov time is related to computational or inherent uncertainties that often lead to overestimating the actual value of the time. To correct this overestimation, Mihailović et al. (2019) introduced the Kolmogorov time as the inverse of the Kolmogorov complexity. This time quantifies the length of the time window within which complexity remains unchanged, significantly while reducing the size of the effective prediction horizon. River regimes can be simple, mixed, or complex, and one question is how these regimes relate to complexity, chaotic behavior, and prediction horizon.

The time horizon of streamflow is a consequence of intertwined hydro-meteorologic forcings (e.g., precipitation, temperature, and evapotranspiration) and physiography (e.g., slope and elevation) (Knoben et al., 2018; Mathai and Mujumdar, 2022). Higher elevation can impact hydro-meteorological dynamics due to more rapid changes in airflow and orographical effects on precipitation production (Houze, 2012). Slope affects the recession of hydrograph (Mathai and Mujumdar, 2022). There is no doubt that these factors (separately or in synergy) may affect the time horizon of rivers. Additionally, the naturalized streamflow data minimize human impacts. However, dam effects may have a dominant influence on the predictability of rivers. It is somewhat unusual that little attention has been paid to this influence in the hydrological literature.

We are of the opinion that the question of streamflow complexity must be approached through information measures to obtain more natural and reliable information. To do that we applied the KC complexity and its derivatives and the Lyapunov exponent to monthly river flow time series over a 66-year period 1950–2015 from 1879 rivers in the United States. Therefore, implications of annual mean precipitation and temperature, slope, elevation, and effects of the existence of dams upstream of the naturalized streamflow site on statistics (i.e., coefficient of variation (CV) and average value) and river time horizon of streamflow were used to explain the results obtained with information measures.

**2. Description of data**

Monthly naturalized streamflow data for a period from 1950 to 2015 was obtained from the U.S. Geological Survey (USGS) Science Base Catalog. The naturalized streamflow is a simulated data for 2,622,273 stream reaches, which are defined by National Hydrography Dataset (NHD) Version 2.0, across the continental U.S. using the random forest ensemble (Miller et al., 2018). However, using all the naturalized data of stream reaches >2.5 million is not only handling the redundant streamflow information but also taking a high likelihood of using less accurate estimates: the random forest models were calibrated to almost 2,000 reference gauge sites where the observed streamflow exists, and then the calibrated models were applied to the ungauged reach segments (Miller et al., 2018). Therefore, we only used the naturalized streamflow data at stream reaches directly connecting the gauge stations at the outlet of HUC (Hydrologic Unit Code) 8 so that naturalized streamflow data from only 1879 sites were used.

    Annual mean precipitation and temperature at every point corresponding to the selected 1879 naturalized streamflow sites were calculated using NOAA nClimGrid monthly dataset. The NOAA nClimGrid dataset has a period from 1895 to the present and covers CONUS and Alaska with a 5 km grid resolution (Vose et al., 2014). Therefore, the nClimGrid precipitation and temperature in the common period for the naturalized streamflow 1950-2015 were spatially interpolated to the naturalized streamflow sites using the inverse distance weight (IDW) method (Ahrens, 2006). Besides, the locations of 92075 dams across CONUS were obtained from the National Inventory Dams (NID). We assigned the binary value of 1 (0) for dams when they are (not) located upstream of the naturalized streamflow site in a watershed for further analysis in the following sections. The mean slope corresponding to each HUC8 watershed was obtained from the U.S. Environmental Protection Agency (EPA), and elevation at each stream gauge point was obtained from the USGS.

The spatial distributions of monthly streamflow, coefficient of variation (CV), altitude, slope, and dams' locations of the U.S. rivers for the period 1950-2015 are shown in Fig. 1.

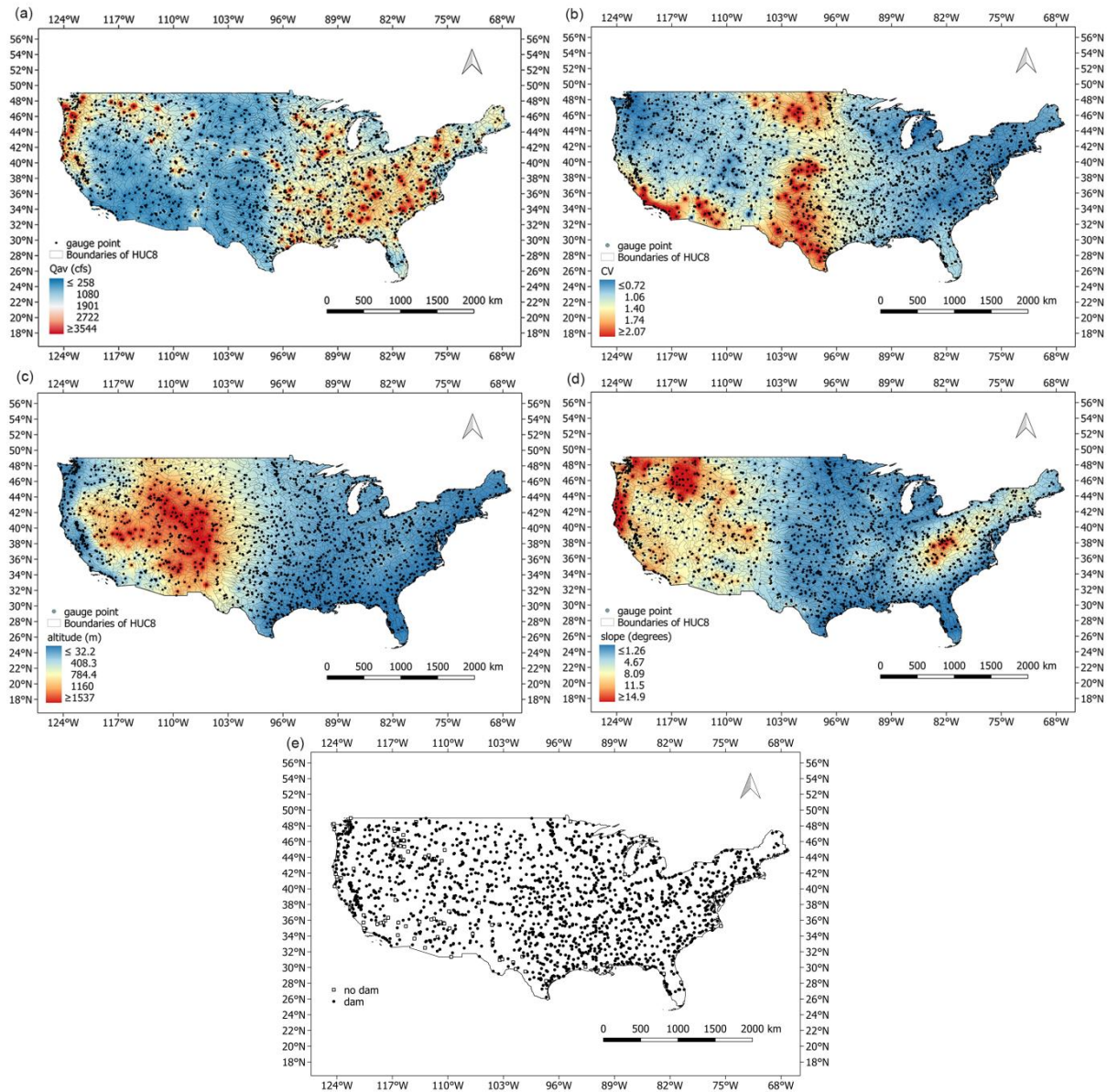

**Fig. 1.** Spatial distributions of (a) maximal streamflow in cfs (cubic feet per second), (b) CV, (c) altitude (m), (d) slope (degree), and (e) dams of the U.S. rivers for the period 1950-2015.

## 3. Methodology

### 3.1 Lempel and Ziv algorithm

Kolmogorov complexity (KC) is a natural but uncomputable information measure. It is approximated by some compression algorithms - Lempel-Ziv and its variants. Lempel and Ziv (1976) suggested an algorithm (LZA) for calculating the complexity of a time series $X(x_1, x_2, x_3, \ldots, x_N)$. It includes the following steps. (1) Encoding the time series by creating a sequence $S$ of the characters 0 and 1 written as $s(i), i = 1, 2, \ldots, N$, according to the rule $s(i) = 0$ if $x_i < x_t$ or 1 if $x_i > x_t$, where $x_t$ is a threshold. The threshold is commonly selected as the mean value of the time series, while other encoding schemes are also available (Radhakrishnan et al., 2000); (2) calculating the complexity counter $c(N)$. $c(N)$ is defined as the minimum number of distinct patterns contained in a given character sequence. The complexity counter $c(N)$ is a function of the length of sequence $N$. The value of $e(N)$ approaches an ultimate value $c(N)$ as $N$ approaches infinity, i.e. $c(N) = O(b(N))$ and $b(N) = log_2 N$; (3). Calculating the normalized information measure $C_k(N)$, which is defined as $C_k(N) = c(N)/b(N) = c(N)/log_2 N$. For a nonlinear time series, $C_k(N)$ varies between 0 and 1, although it can be larger than 1. Note that the pattern is a sequence in the coded time series which is unique and non-repeatable. A flow chart for the calculation of KC of a streamflow series $X(x_1, x_2, x_3, \ldots, x_N)$ using the LZA algorithm is shown in Fig. 2.

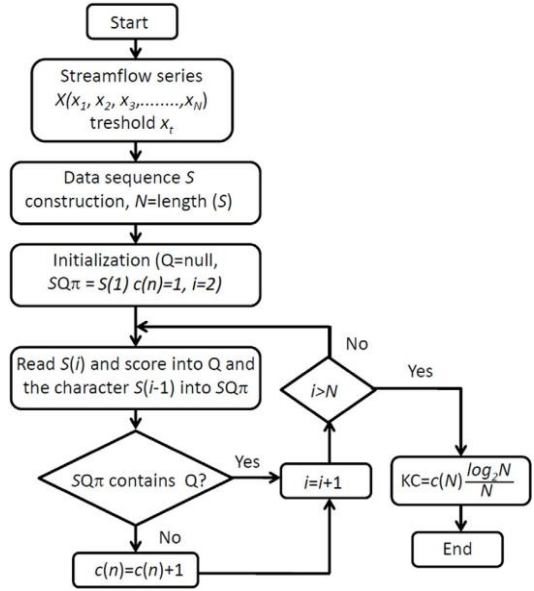

**Fig. 2.** Flow chart for calculation of the Kolmogorov complexity (KC) using the Lempel–Zev algorithm (LZA) (by permission Mihailović et al., 2019)

*3.2 Kolmogorov complexity spectrum and its highest value*

The Kolmogorov complexity of time series has two weaknesses: (i) it cannot distinguish between time series with different amplitude variations and that with similar random components; and (ii) in the conversion of a time series into a binary string, its complexity is unseen in the rules of the applied procedure. Therefore, in defining a threshold for a criterion for coding, some information about the composition of time series could be lost. In the complexity analysis of time series, two measures are used: (i) Kolmogorov complexity spectrum (KC spectrum) and (ii) the highest value of KC spectrum (KCM), introduced by Mihailović et al. (2015a) who described the procedure for calculating the KC spectrum. The flow chart in Fig. 3 shows schematically how to calculate the KC spectrum $C(c_1, c_2, c_3, \ldots, c_N)$ for time series $X(x_1, x_2, x_3, \ldots, x_N)$. This spectrum allows us to investigate the range of amplitudes in a time series that represents a complex system with highly enhanced stochastic components. It may be noted that for a large number of samples of a time series, the computation of KC spectrum can be challenging. The highest value $K_m^C$ as in this series, i.e., $K_m^C = max\{c_i\}$, is the highest value of the KC complexity spectrum.

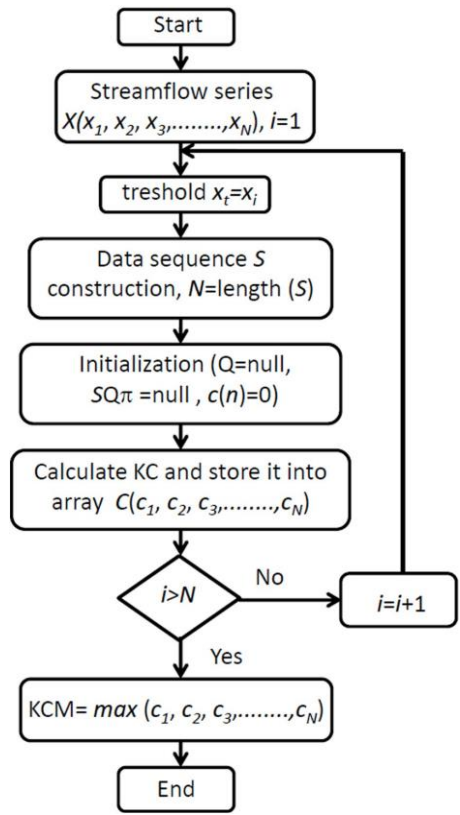

**Fig. 3.** Flow chart for calculation of the Kolmogorov complexity spectrum and its highest value (KCM) (by permission Mihailović et al., 2019).

*3.3 Lyapunov exponent*

The Lyapunov exponent of a dynamical system is a quantity that characterizes the rate of separation of infinitesimally close trajectories. Positive Lyapunov exponent (LE) indicates that small fluctuations can lead to drastically different system behavior (small differences in the initial state lead to large differences in a later state). Because the rate of separation can be different for different orientations of the initial separation vector, there is a spectrum of Lyapunov exponents whose largest value is commonly the LE. A positive value of this exponent is taken as an indicator that a dynamical system is chaotic. In this study, we obtained the LE for the standardized monthly streamflow time series by applying the Rosenstein algorithm (Rosenstein et al., 1993) which was implemented in the MATLAB program (Shapour, 2009). However, this measure has one drawback: If the embedding theory is used to build chaotic attractors in the reconstruction space, then additional "spurious" Lyapunov exponents appear.

4. **Results and Discussion**

*4.1 Spatial analysis*

*4.1.1 Picture of the US rivers: complexity and chaos*

The scatter plot of KC versus LE is like a "boomerang" shape (Fig. 4). The scatter plot area, by two lines LE = 0.146 (parallel to the y-axis) and KC=0.516 (parallel to the x-axis), is divided into rectangles for the sake of a better visualization for further analysis. These two numbers are the means of the maximal and minimal values of KC and LE in the set of gauge stations. Perhaps, at first glimpse, this kind of scatter plot (with a large number of samples) seems surprising and may appear for the first time to our knowledge. The figure shows that streamflow time series for all gauge stations give a picture of a mixture consisting of always present chaos and high randomness. Randomness is unpredictable because we just have no right information, while chaos is somewhere between random and predictable. A hallmark of chaotic streamflow time series is predictability. Thus, with this picture of the scatter plot of KC versus LE, it can be said that it is difficult to reach a reliable prediction of the U.S. river streamflow. In addition, depending on the time scale considered, river discharge can be either random or chaotic. Thus, at daily and seasonal scales the river discharge is random (non-chaotic), but the flow is chaotic at the monthly scale (Adab et al., 2018). A simple percentage calculation with numbers from Fig. 4 shows that the LU values of all gauge stations are positive, while 83.7 % of their KC values are higher than 0.516 (LU and RU parts). In the two lower parts (LD and RD) randomness and LE of streamflow are mainly lower. From Fig.4, it is simple to find that the ratio of the number of gauge stations with low and high randomness is approximately one to five. This is an indicator that the predictability of the U.S. river streamflow is not high, i.e., we talk about the chance that river streamflow could be in principle modeled reliably. We do not talk about heuristic hydrological models, which sometimes can give good results that come from their mathematical background.

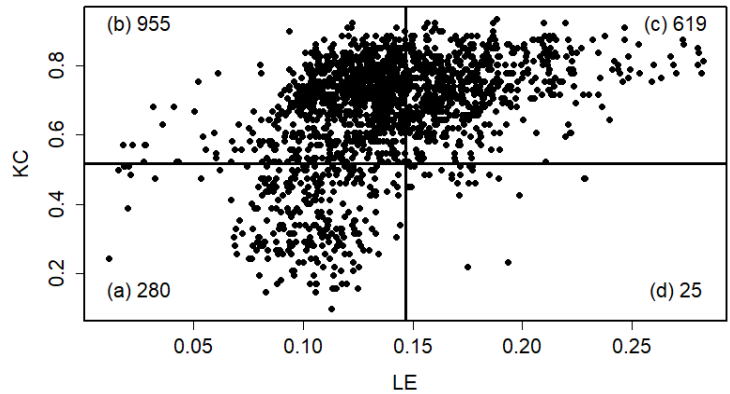

**Fig. 4.** Scatter plot of Kolmogorov complexity (KC) versus, and Lyapunov exponent (LE) for the U.S. monthly streamflow for the period 1950-2015. (a) LD (left-down); (b) LU (left-upper); (c) RU (right-upper) and (d) RD (right-down) are parts of the scatter plot area. The numbers in rectangles indicate the total number of gauge stations.

*4.1.2 Complexity*

The complexity of river discharge is a key issue in hydrology. This paper considers the KC complexity. The spatial distribution of KC of the US rivers' monthly streamflow for the period 1950-2015 is shown in Fig. 5a. From this figure three distinct patterns for KC are seen, and more if we consider a finer scale. The southwestern part which is arider has lower KC than what is in the more humid part of the eastern U.S. Further, in the southwestern part, there is a band of very low KC. This band cuts across the mountainous terrain. The Mississippi valley, the Great Lakes region, and the Atlantic Seaboard have high KC. However, the Ohio Valley has a lower (but still high) KC than the surrounding area. It seems that besides landscape and physiographic characteristics, KC may have a strong correlation with weather patterns and distance from the sea (continentality). One can make a similar observation about LE (Fig. 5b). It is quite difficult and not always consistent to determine explicitly the connection between physiographic characteristics and the complexity of river flow. For example, based on monthly streamflow time series from ten-gauge stations at seven rivers of different river regimes in Bosnia and Herzegovina, Mihailović et al. (2015b) found that the relationship between the highest value of the KC spectrum (KCM) and elevation (h). That relation KCM=0.0002h + 0.9421 shows that there is a positive trend in changes of the KCM with respect to elevation with the coefficient of correlation of 0.602

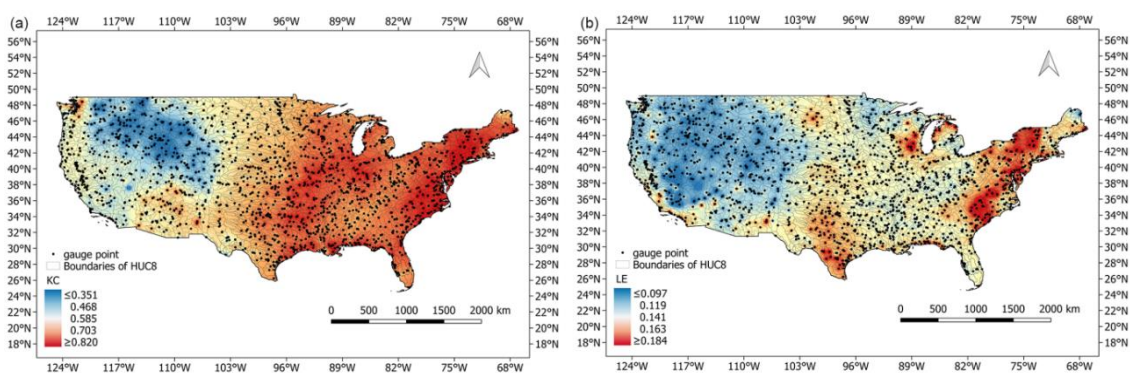

**Fig. 5.** (a) Spatial distribution of KC and (b) LE of the U.S. river monthly streamflow for the period 1950-2015.

We hypothesize that the turbulent nature of rivers and barriers built by human activities may remarkably affect the complexity of rivers. When we say human activities, we primarily mean dams but also channels. Figure 1e shows the spatial distribution of dams (1796) built on the U.S. rivers making up 95.6% of the statistical set of 1879 gauge stations used in this study. A small number of papers are devoted to this issue. And if there are any, they mostly remain on the descriptive approach supported by traditional statistics. To address the complexity of streamflow, Mihailović et al. (2019) analyzed daily streamflow data recorded during the period 1989–2016 at twelve gauging stations on the Rio Brazos River in Texas (USA) using KC and its derivatives. They found a huge increase in KCM at one gauge station in comparison to other ones, concluding that the reason for the high KCM of this station may be attributed to the Morris Sheppard Hydroelectric Power plant at Morris Sheppard Dam. The KC complexity as a measure does not distinguish between time series with different amplitude variations and similar random components. It seems that changing the river flow in the operating mode of the dam does not only change the amplitude but also the randomness, which is reflected by higher complexity, i.e., what is captured by KC appropriately. In this paper, the range of the calculated measures was in the intervals of 0.097-0.936 (KC) and 0.011-0.282 (LE), respectively. The spatial distribution of KC in Fig. 6a is very similar to the distribution of KC in Fig. 5a. In Fig. 6a the number of dams is 1796, while the number of dams among gauge stations with KC > 0.516 is 1523 or 84.7% of the total number of dams. This is almost the same as the ratio of 83.7% (number of gauge stations with KC > 0.516 and their total number) in 4.1.1.

It is noted that the channelization of rivers decreases the complexity of river flow. For example, applying the KC complexity analysis of monthly river discharge, Mihailović et al. (2014) found that during the period 1926–1990 there was a drop in KC in the mountain rivers in Bosnia and Miljacka (Bosnia and Herzegovina) for the period 1946-1965, in comparison with two other periods (1926-1945 and 1966-1990). That complexity loss was interpreted as a result of intensive different human interventions on those rivers (establishing the network of channels for building the capacities for water consumption) after the Second World War. Certainly, channelization is a type of human intervention that contributes to reducing the randomness of the U.S. rivers having KC less than 0.516 with an amount of 16.2% of the total number of gauge stations. The division of river flow regimes into (i) low complex (KC < 0.516) and high complex (KC ≥ 0.516) and (ii) low chaotic (LE < 0.146) and high chaotic (LE ≥ 0.146) is merely conditional.

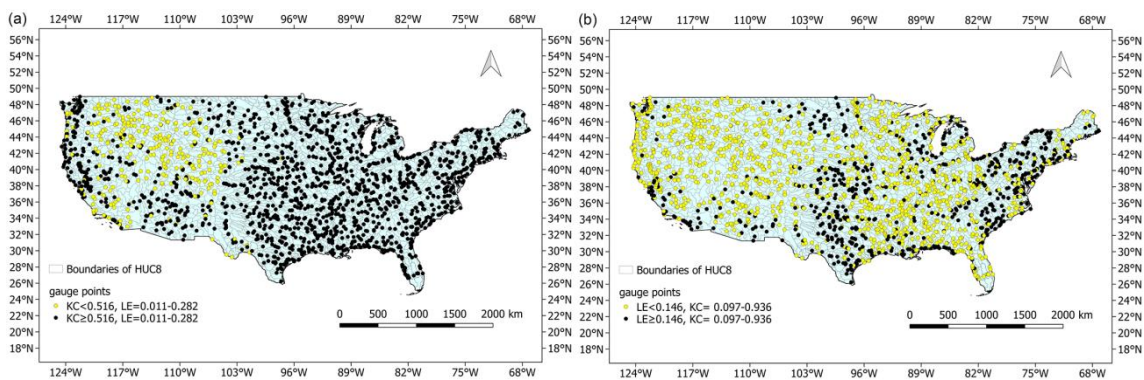

**Fig. 6**. (a) Spatial distribution of KC (KC < 0.516; KC ≥ 0.516 while 0.011 < LE < 0.282) and (b) LE (LE < 0.146; LE ≥ 0.146 while 0.097 < KC < 0.936) of the U.S. river monthly streamflow in the presence of dams for the period 1950-2015.

*4.1.3 Chaotic behavior*

The spatial distribution of LE of monthly discharge of the U.S. rivers for the period 1950-2015 is shown in Fig. 6b. This is the spatial distribution taken out from the distribution in Fig. 5b following the mentioned criterion of division of U.S. river flows into low chaotic (LE < 0.146) and high chaotic (LE ≥ 0.146). From this figure it is seen that low chaos prevails significantly. High chaos dominates in the Mississippi Valley, the Great Lakes region, the Upper Mid-West, and the Atlantic Seaboard, while in the Western U.S. it is less prevalent.

We already stated that chaos is always present in turbulent flow. Therefore, the rivers, which have positive LE, are in a chaotic regime. This phenomenon is intriguing as a topic in hydrology as well as in all sciences. However, hydrologists mostly have dealt with those topics in the following way. (1) They often keep on a descriptive level that is not always necessarily simple, paying more attention to the mathematical background with comments about possible applications. (2) They usually consider just a low-dimensional chaos, i.e., to be attributable to a small fraction of the components of the total system, using a smaller number of streamflow time series with a focus on one possible source that causes that chaos. (3) Some of them used theoretical approaches, inverse modeling, and information measures to gain insights into the river flow's chaotic nature. (4) There is almost no information that anyone dealt with the high-dimensional chaotic regime of river flow, i.e., turbulent flow having many degrees of freedom (Porporato and Ridolfi, 1997; Sivakumar, 2000; Sivakumar and Jayawardena, 2002; Labat et al., 2011; Fattahi et al., 2013; Yildirim et al., 2016; Mihailović et al., 2019).

The spatial distribution of LE (0.09-0.18) of monthly streamflow of the U.S. rivers for the period 1950-2015 is shown in Fig. 7a. This time interval was chosen so that their endpoints were nearly symmetrical with respect to LE = 0.146. The spatial distribution of gauge stations in this figure in comparison with the spatial distribution of dams in Fig. 1e is almost indistinguishable. Further inspection of Fig. 7a shows that the number of gauge stations in the Western part (which is mostly mountainous) is approximately the same as the number of gauge stations in the Eastern part. This leads us to the conclusion that the influence of orography on the river flow dynamics is much smaller than the influence of dams, i.e., the influence of human activity.

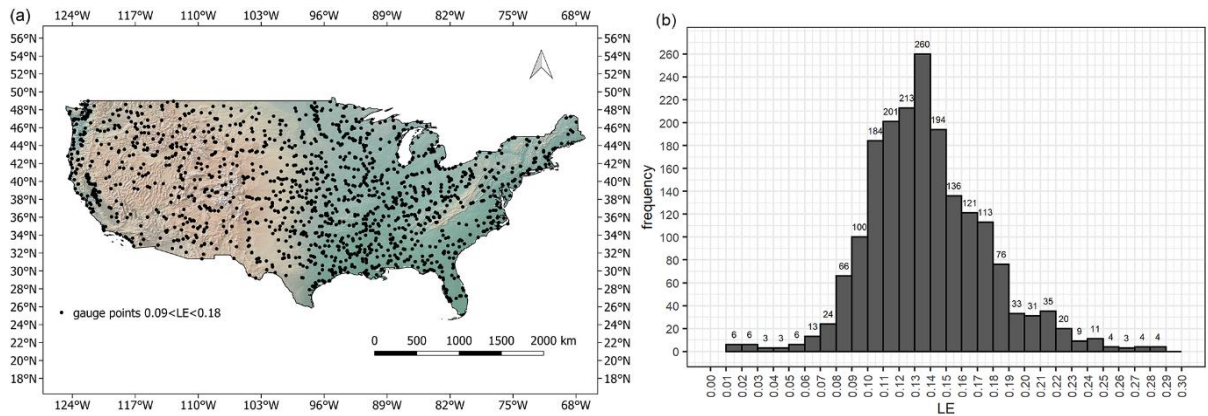

**Fig. 7.** (a) Spatial distribution of river gauge stations (0.09 < LE < 0.18) of monthly discharge of the U.S. rivers for the period 1950-2015 and (b) histogram of their numbers in dependence on LE.

If we look at the scatter plot (Fig. 4) we can see that LE values of gauge stations from Fig. 7a are mostly grouped on the right side of LU and the left side of RU. The number of gauge stations with a frequency greater or equal to 100 is 1522 (Fig. 7b), or 81.0% of the total number of stations. It is interesting to consider the "tail" of the histogram in Fig. 7b. It corresponds to the state of high chaos and low complexity of river flow (RD part in Fig. 4). This may indicate the appearance of periodicity or another pattern on some other time scales that can be ascribed to specific environmental factors or human activity (Aksentijevic et al., 2020). The occurrence of periodicity on other time scales does not necessarily mean that it will be maintained over long periods.

*4.2 Time horizon and complexity spectrum of river flow amplitudes*

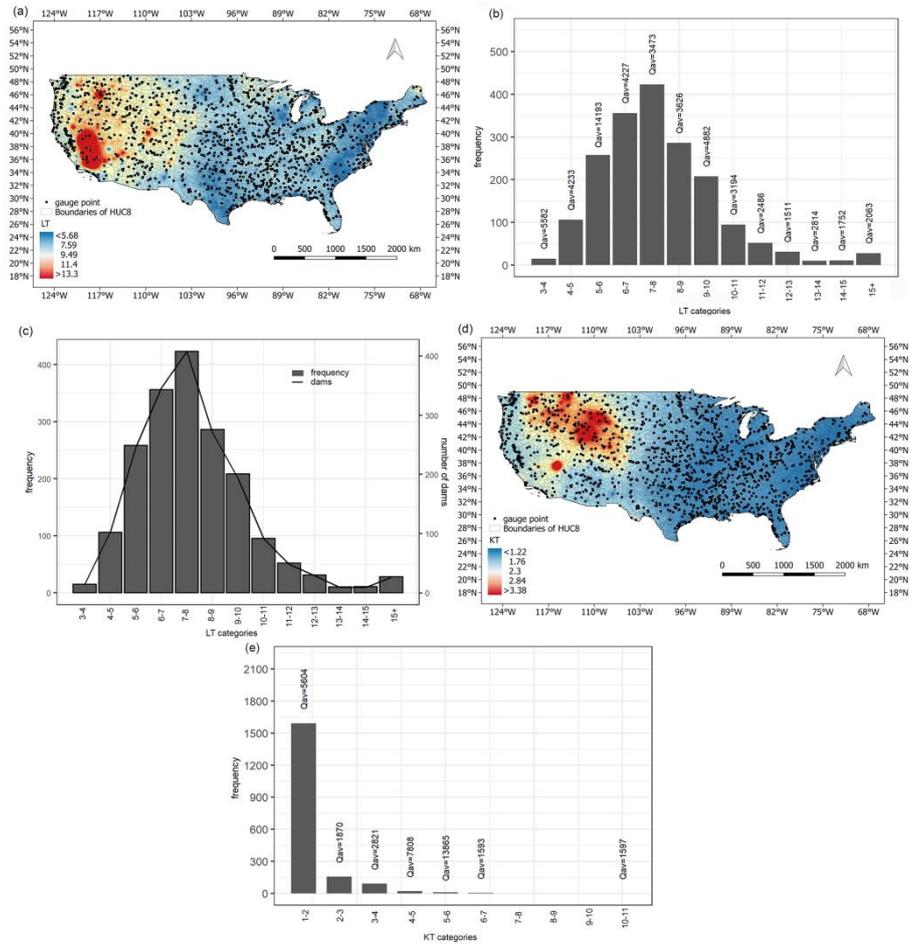

**Fig. 8.** (a) Spatial distribution of time horizon of the U.S. rivers (LT is in moths); (b) histogram of LE; (c) distribution of dams versus LE frequency; (d) spatial distribution of KT for the U.S. rivers (KT is in moths) and (c) histogram of KT.

The Lyapunov time (LT) reflects the limits of predictability of the dynamical system, while the Kolmogorov time (KT) indicates the limitations in predictability due to the presence of randomness. Figure 8a shows the spatial distribution of LT for the U.S. rivers. Coarsely, LT can be classified as follows. (1) LT < 6 (Southeast region, parts of the Southwest region, and isolated island in the northern part of the Midwest region); lower predictability of rivers may be primarily attributed to the presence of dams in synergy with the type of orography. (2) Areas with LT > 6 can be observed especially in the Great Lakes area. (3) In the western region and part of the Southwest, the LT values are high and in some parts they exceed the value of 13. The lower LT values include a narrow strip immediately along the coast. If the spatial distribution of LT is transferred to a histogram (Fig. 8b), then the causes of this spatial distribution become more understandable. The histogram shows that in the LT categorization, the largest number of gauge stations is located in the interval (5-6) – (9-10). Such LT values are the result of the LE distribution in Figure 7b, where the frequency distribution is the highest in the interval (0.09, 018). The histogram in Figure 7c shows that the largest number of the U.S. rivers have LE values that are in the interval between 0.125 and 0.143, i.e., with a time horizon that is between seven and eight months.

In this paper, we try to see which factor has the greatest influence on the time horizon of rivers. It seems to be a factor that affects the dynamics of river flow originating from human activities. Certainly, there are other factors, but this one is extremely dominant, as can be seen in Fig. 8c. It shows a high correspondence between LT and the number of dams on rivers. This was pointed out by Mihailović et al. (2019) where LE, as well as KC, were analyzed at twelve stations on the Brazos River. The influence of dams on river flow is discussed in subchapter 4.1.2 where the influence on KC is considered. That influence (changes in river flow operating and nonoperating mode of dams) reflects also on LE, i.e., on LT. Figures 8d and 8e show the spatial distribution and histogram of KT of the U.S. rivers from which it is seen that that time is no longer than 1-2 months. That time quantifies the time window size within which complexity remains unchanged. Hence, the presence of a narrow window significantly reduces the length of effective prediction horizon. Thus, the relationship between KT, LT, and Qav may provide a deeper insight into the predictability of river streamflow under the KC spectrum of mean monthly streamflow (Qav), for different LT categories. To obtain the KC spectrum for LT categories (5-9 months), a time series of mean monthly streamflow was formed by averaging over all gauge stations whose LT was in those LT categories. These LT categories were selected using Fig. 8b. The KC spectrum was determined for that time series. The number of stations was: (1) Qav < 500 cfs (38.1 % of the total number of gauge stations (1531 with 1465 dams) in the considered interval of LT); (2) 500 ≤ Qav < 2500 cfs (36.2%); (3) 2500 ≤ Qav < 8500 cfs (16.1%) and (4) Qav ≥ 8500 cfs (9.6%-gauge stations). From Fig. 9 is seen that the dependence of zones of KC on mean monthly streamflow Qav can be distinguished as follows.

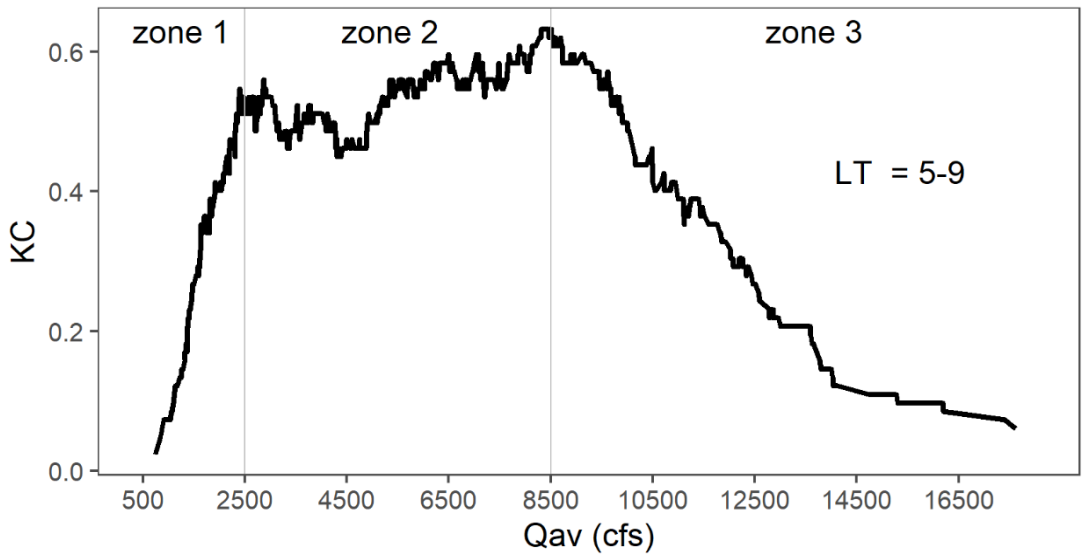

**Fig. 9.** Kolmogorov complexity spectrum of mean monthly streamflow Qav (cfs) of the U.S. rivers for different categories (5-9 months); the number indicates the left side point of one month's interval) of the Lyapunov time (LT). Streamflow zones are divided by vertical lines.

*Zone* 1 (500 ≤ Qav < 2500 cfs) is a zone of increasing KC and LE (ring of gauge stations in the West, central and eastern parts of the Midwest, the eastern part of the Southwest, Southeast, and Northeast, as shown in Fig.10a). In the course of rivers, randomness is more prevalent, which, despite being high (KT is lower), does not significantly affect the time horizon, since the smaller values of LE (around 0.139 on average for the LT categories (5-9 months); see Fig. 8b) increase the predictability. *Zone* 2 (2500 ≤ Qav < 8500 cfs). This zone has a distribution that is similar to zone 1. In its part of the KC spectrum, the fluctuations in complexity are emphasized moving towards higher streamflows, but in an increasing trend. The comment for the time horizon in this zone is similar to zone 1, with the predictability increase going towards higher river streamflow. *Zone* 3 (Qav ≥ 8500 cfs). This zone includes the smallest number of stations placed in the northern part of the West, partly in the Southwest and eastern part of the U.S.. In this zone, KC, as well as LE, decreases resulting in a long time of predictability.

Qav < 500 cfs is a zone of very low KC and also lower LE (38.1% gauge stations). This zone includes a band extending along the western part of the Midwest and the eastern part of the Southwest, the central part of the West, and a narrow belt along the Atlantic coast in the Northeast (Fig. 10a). Low values of KC and LE result in a longer time horizon. This zone is not visible on the KC spectrum due to extremely small KC values and also due to averaging, but it exists.

Figure 10b visualizes three-dimensional relationship between KT, LT, and Qav, which includes LT time horizon categories of 5-9 months.

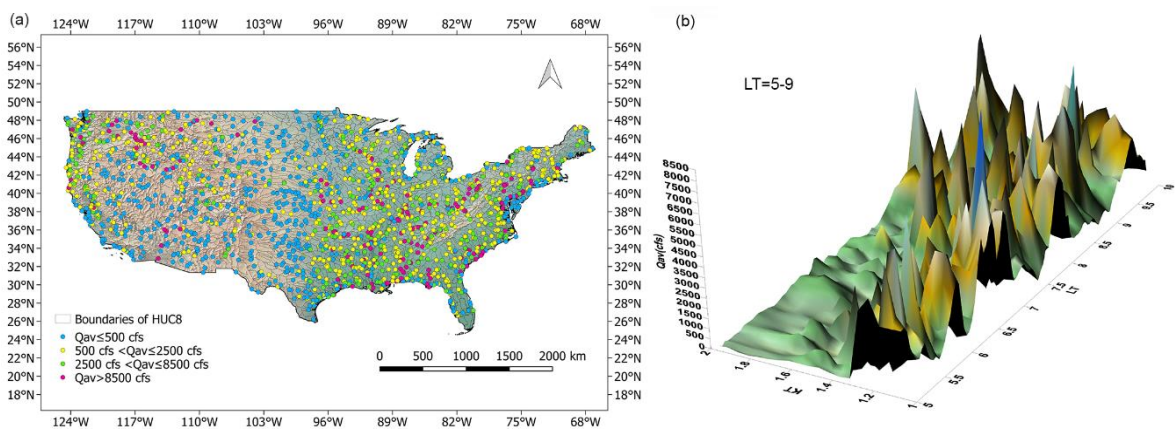

**Fig. 10.** The map mean LT time horizon categories (5-9 months) for different streamflow intervals of the U.S. rivers and (b) three-dimensional visualization of mean KT, LT versus streamflow (Qav).

Undoubtedly, dams also and other types of human activity affect the dynamics of rivers. However, many other factors, such as weather, climate, orography, continentality of the place, etc., separately or in synergy, affect the size of time horizon, i.e. predictability of streamflow. It seems that the influence of these factors is less significant than is human activity. Although the focus of this paper is not the quantification of influence of environmental and other natural factors on the LT time horizon categories of rivers, we list some averages of those factors. It is done for gauge stations having the LT time horizon categories (5-9), i.e. the longest ones for the U.S. rivers. They are (1) the number of gauge stations (1531); (2) the number of dams (1465); monthly streamflow (5676 cfs); CV (1.207); slope (5.1 degrees); altitude (400 m); temperature (12.7 $^{o}$C); and average annual precipitation (33.3 inches).

## 5. Conclusions

Monthly streamflow data (1950-2015) from 1879 gauge stations on the U.S. rivers were analyzed using the Kolmogorov complexity (KC) and related complexity measures (Kolmogorov complexity spectrum) and Lyapunov exponent (LE) to establish the time horizon of rivers by calculating the Lyapunov time (LT) and Kolmogorov time (KT). The following conclusions are drawn from this study:

(1) The values of calculated measures were in the intervals 0.097-0.936 (KC) and 0.011-0.282 (LE), respectively;

(2) The number of gauge stations with KC > 0.516 was 1574 (83.7% of the total number of gauge stations (1879) while LE > 0 was obtained for all gauge stations;

(3) The high complexity (KC) and the presence of chaos in the streamflow of all U.S. rivers (LE is always positive) may be addressed to human activities, primarily in the presence of a large number of dams (1796 or 95.6% of the total number of gauge stations); with their mode of operation they introduce significant changes in the complexity and the turbulent flow of rivers, increasing the level of chaos);

(4) The West region and southwestern part of the Southwest region have LT (Lyapunov time or time horizon) between 10 and 13 (or more) months; the rest of the Southwest region, Midwest, Southeast, and Northeast regions have LT between 5 and 9 months;

(5) The number of gauge stations with LT between 5 and 9 months is 1531 with the following frequency distribution in relation to LT categories (in months): 258 (5-6), 356 (6-7), 423 (7-8), 286 (8-9), and 208 (9-10);

(6) Human activity affects the dynamics of rivers but many other factors, such as weather, climate, orography, continentality of the place, etc., separately or in synergy, affect the size of time horizon that was not the focus of the paper. However, there is a justified expectation that the connection of Kolmogorov complexity and Lyapunov exponent with environmental factors can quantify their role in the predictability of streamflow without the use of traditional mathematical statistics. It will be the content of our forthcoming paper.

**Declaration of Competing Interest**

The authors declare that they have no known competing financial interests or personal relationships that could have appeared to influence the work reported in this paper.

**Data availability**

The data authors used are publicly available online: Monthly naturalized streamflow data (at https://www.sciencebase.gov/catalog/item/59cbbd61e4b017cf314244e1), NOAA nClimGrid monthly precipitation and temperature data (at https://www.ncei.noaa.gov/access/metadata/landing-page/bin/iso?id=gov.noaa.ncdc:C00332),

National Inventory Dams data (at https://www.fema.gov/emergency-managers/risk-management/dam-safety/national-inventory-dams), The mean slope of watershed from EPA (at https://www.epa.gov/wsio/wsio-indicator-data-library), Elevation of gauge station from USGS (at https://waterdata.usgs.gov/nwis/sw).